\documentclass[twocolumn]{aastex702}

\usepackage{comment}
\usepackage{booktabs}

\begin{document}
\title{IGRINS-2 Reveals an Oxygen-rich Dayside Atmosphere of WASP-12~b: Resolving the C/O Controversy}

\author[0000-0001-6876-8559]{Changwoo Kye}
\affiliation{Department of Physics and Astronomy, Seoul National University, Seoul 08826, Republic of Korea}
\affiliation{Korea Astronomy and Space Science Institute, 776 Daedeok-daero, Yuseong-gu, Daejeon 34055, Republic of Korea}
\email[show]{kyecw1@gmail.com}

\author[0009-0005-5145-5165]{Yeon-Ho Choi}
\affiliation{Korea Astronomy and Space Science Institute, 776 Daedeok-daero, Yuseong-gu, Daejeon 34055, Republic of Korea}
\email{yhc@kasi.re.kr}

\author[0009-0005-4890-3326]{Krishna Kanumalla}
\affiliation{School of Earth and Space Exploration, Arizona State University, Tempe, AZ 85287, USA}
\email{skanuma1@asu.edu}

\author[0000-0002-2338-476X]{Michael Line}
\affiliation{School of Earth and Space Exploration, Arizona State University, Tempe, AZ 85287, USA}
\email{mrline@asu.edu}

\author[0000-0003-0894-7824]{Jae-Joon Lee}
\affiliation{Korea Astronomy and Space Science Institute, 776 Daedeok-daero, Yuseong-gu, Daejeon 34055, Republic of Korea}
\email{lee.j.joon@gmail.com}

\author[0000-0002-8573-805X]{Stefan Pelletier}
\affiliation{Department of Physics and Trottier Institute for Research on Exoplanets, Universit\'{e} de Montr\'{e}al, 1375 Avenue Thérèse-Lavoie-Roux, Montreal, H2V 0B3, Canada}
\email{stefan.pelletier@umontreal.ca}

\author[0000-0002-9946-5259]{Peter Smith}
\affiliation{School of Earth and Space Exploration, Arizona State University, Tempe, AZ 85287, USA}
\affiliation{Max-Planck-Institut f\"ur Astronomie, Heidelberg 69117, Germany}
\email{petercbsmith@asu.edu}

\author[0000-0002-2513-4465]{Vatsal Panwar}
\affiliation{School of Physics \& Astronomy, University of Birmingham, Edgbaston, Birmingham B15 2TT, UK}
\email{panvatsal@gmail.com}

\author[0000-0001-8877-0242,sname='Hayashi']{Yuya Hayashi}
\affiliation{Department of Multi-Disciplinary Sciences, Graduate School of Arts and Sciences, The University of Tokyo, 3-8-1 Komaba, Meguro, Tokyo 153-8902, Japan}
\email{yuya-hayashi@g.ecc.u-tokyo.ac.jp}

\author[0000-0001-8969-0009]{Jae-Rim Koo}
\affiliation{Korea Astronomy and Space Science Institute, 776 Daedeok-daero, Yuseong-gu, Daejeon 34055, Republic of Korea}
\email{koojr@kasi.re.kr}

\author[0000-0003-0815-8366]{Swaetha Ramkumar}
\affiliation{School of Earth and Space Exploration, Arizona State University, Tempe, AZ 85287, USA}
\email{sramku14@asu.edu}

\author[0000-0002-5847-8096]{Sung-Chul Yoon}
\email{scyoon@snu.ac.kr}
\affiliation{Department of Physics and Astronomy, Seoul National University, Seoul 08826, Republic of Korea}

\begin{abstract}
Recent studies have shown that measurements of hot-Jupiter atmospheric composition, including carbon-to-oxygen (C/O) and refractory-to-volatile (R/V) abundance ratios, provide powerful diagnostics of the chemistry and formation of gas giant planets.
Important questions remain about the atmospheric composition of the ultra-hot Jupiter WASP-12~b: its dayside C/O ratio remains under debate, and its R/V ratio is not yet well constrained.
We present high-resolution (R $\sim$ 45,000) near-infrared dayside emission spectroscopy of the ultra-hot Jupiter WASP-12~b with IGRINS-2 on Gemini North during post-eclipse phases.
Cross-correlation analysis reveals volatile molecular signals of H$_2$O and CO together with tentative signals of OH and gas-phase refractories in the planetary atmosphere.
The planetary signals appear in emission, indicating a thermal inversion in the dayside atmosphere.
A chemically consistent Bayesian retrieval yields C/O $=0.72^{+0.08}_{-0.10}$, disfavoring carbon-rich dayside solutions with C/O $>1$ at greater than $5\sigma$.
This independent, line-resolved measurement supports an oxygen-rich atmosphere, providing a new constraint on the long-standing C/O controversy surrounding WASP-12~b.
Our retrieval also yields a host-star-normalized R/V ratio, $[\mathrm{R/V}]_{*}=0.17^{+0.22}_{-0.23}$, which is consistent with those of other ultra-hot Jupiters similarly characterized with IGRINS/IGRINS-2.
Although the sample remains small, this consistency suggests a shared atmospheric chemistry among these planets, potentially reflecting similar proportions of accreted volatile and refractory material.
\end{abstract}


\section{Introduction} \label{sec:intro}
Understanding how hot Jupiters form remains a central topic in exoplanet science. Atmospheric composition provides one of the few observational records of the gas and solids accreted during planet formation.
In particular, the atmospheric C/O ratio has been widely used to trace formation location because the gas-phase abundances of carbon and oxygen vary across the major snow lines of protoplanetary disks \citep{oberg11}.
Also, the C/O ratio is important for determining atmospheric chemistry.
The abundances of O-bearing molecules can vary with the C/O ratio because each molecule has a different binding energy \citep{madhu12}.
However, C/O alone does not uniquely distinguish among formation histories. A superstellar C/O ratio may result from formation beyond the H$_2$O snow line, but it can also arise closer to the host star if a substantial fraction of oxygen is sequestered in silicates and rocky material \citep{mordasini16}.
Therefore, measurements of both volatile and refractory abundances provide additional leverage for distinguishing among possible formation locations and accretion histories \citep{lothringer21,schneider21a, schneider21b,turrini21,chachan23}.

Ultra-hot Jupiters (UHJs) are particularly favorable targets for applying this joint volatile--refractory diagnostic.
Their high dayside temperatures allow many refractory species to remain in the gas phase, while their strong thermal emission facilitates high-resolution spectroscopy.
By resolving individual spectral lines and exploiting the planet's changing orbital velocity, high-resolution spectroscopy can separate the planetary spectrum from stellar and telluric signals \citep{snellen25}.
Recent near-infrared observations have demonstrated sensitivity to volatile molecules such as H$_2$O, CO, and OH, as well as refractory atoms, in UHJ atmospheres \citep{line21,bro23,ramkumar23,smith24b,choi25,sanchez26,kanumalla26}.
In particular, the simultaneous H- and K-band coverage of IGRINS-2 enables volatile and refractory tracers to be examined within a single data set, reducing potential cross-instrument and temporal systematics in their relative abundance measurements \citep[e.g.,][]{smith24b, pelletier25}.

WASP-12~b is an especially compelling target for such an investigation.
This archetypal UHJ exhibits strong thermal emission and has been extensively characterized at optical and infrared wavelengths.
Early analyses of its dayside atmosphere favored a carbon-rich atmosphere with C/O $\geq1$ \citep{madhu11,stevenson14}, whereas \citet{line14} obtained a dayside solution with C/O $<1$.
HST/WFC3 transmission spectroscopy also favored C/O $<1$ at the terminator \citep{kreidberg15}, while subsequent dayside retrievals and reanalyses have continued to yield contrasting interpretations \citep{oreshenko17,himes22,akinsanmi24}.
These results motivate an independent measurement of the dayside composition.
Although space-based telescopes can probe both volatile and refractory species \citep{gapp25, pelletier26}, high-resolution near-infrared spectroscopy offers complementary constraints on WASP-12~b's C/O ratio and relative volatile and refractory abundances.
In this letter, we present post-eclipse dayside observations of WASP-12~b obtained with Gemini North/IGRINS-2. 
We use cross-correlation analysis and atmospheric retrievals to search for molecular and atomic species and constrain the planet's dayside thermal structure and elemental composition. We then compare the inferred host-star-relative volatile and refractory abundances with abundance measurements for other UHJ observations and predictions from disk-chemistry models to examine their implications for the formation of WASP-12~b.

\section{Observations and Data Reduction} \label{sec:obs}
We observed the dayside thermal emission of WASP-12~b with Gemini North/IGRINS-2 on 2025 February 25 as part of program GN-2025A-Q-112. IGRINS-2 provides simultaneous wavelength coverage from 1.45 to 2.45 $\mu$m at a resolving power of $R\sim45{,}000$. We obtained a continuous 4.5 hr time series comprising 41 A--B nodding pairs (82 individual exposures of 150 s each).
The sequence covered post-eclipse orbital phases $0.561<\phi<0.732$, calculated using the ephemeris of \citet{kokori23}.
During the observation, the airmass increased from approximately 1.0 to 2.1, and the median S/N per A--B pair was approximately 200 in the H band and 150 in the K band.

The raw frames underwent standard extraction and calibration with version 3.0 of the IGRINS Pipeline Package \citep{plp, plp3}, followed by custom processing based on the \texttt{IGRINS-transit} package \citep{mansfield24} adapted for time series observations. 
We discarded orders severely affected by telluric absorption, trimmed 100 pixels from both ends of each retained order, and performed secondary wavelength calibration to correct sub-pixel shifts by aligning sky emission lines in each target frame to those in the blank-sky reference frame using a linear shift-and-stretch transformation.
Quasi-stationary stellar and telluric contamination was identified and removed by applying singular value decomposition (SVD) independently to each spectral order. 
To optimize the number of singular components, we adopted the differential optimization method described by \citet{cheverall23}.
We tested the removal of two to fifteen components and adopted seven components because this choice maximized the S/N of the $\Delta$CCF at the expected planetary velocity.
The results of this test are presented in Appendix \ref{appendix:SVD}.
The resulting residual spectra were used in the cross-correlation analysis described in Section~\ref{sec:method}.

\section{Analysis and Results}
\subsection{Spectrum Templates and Cross-correlation Analysis}
\label{sec:method}

\begin{figure*}
\epsscale{1.2}
\plotone{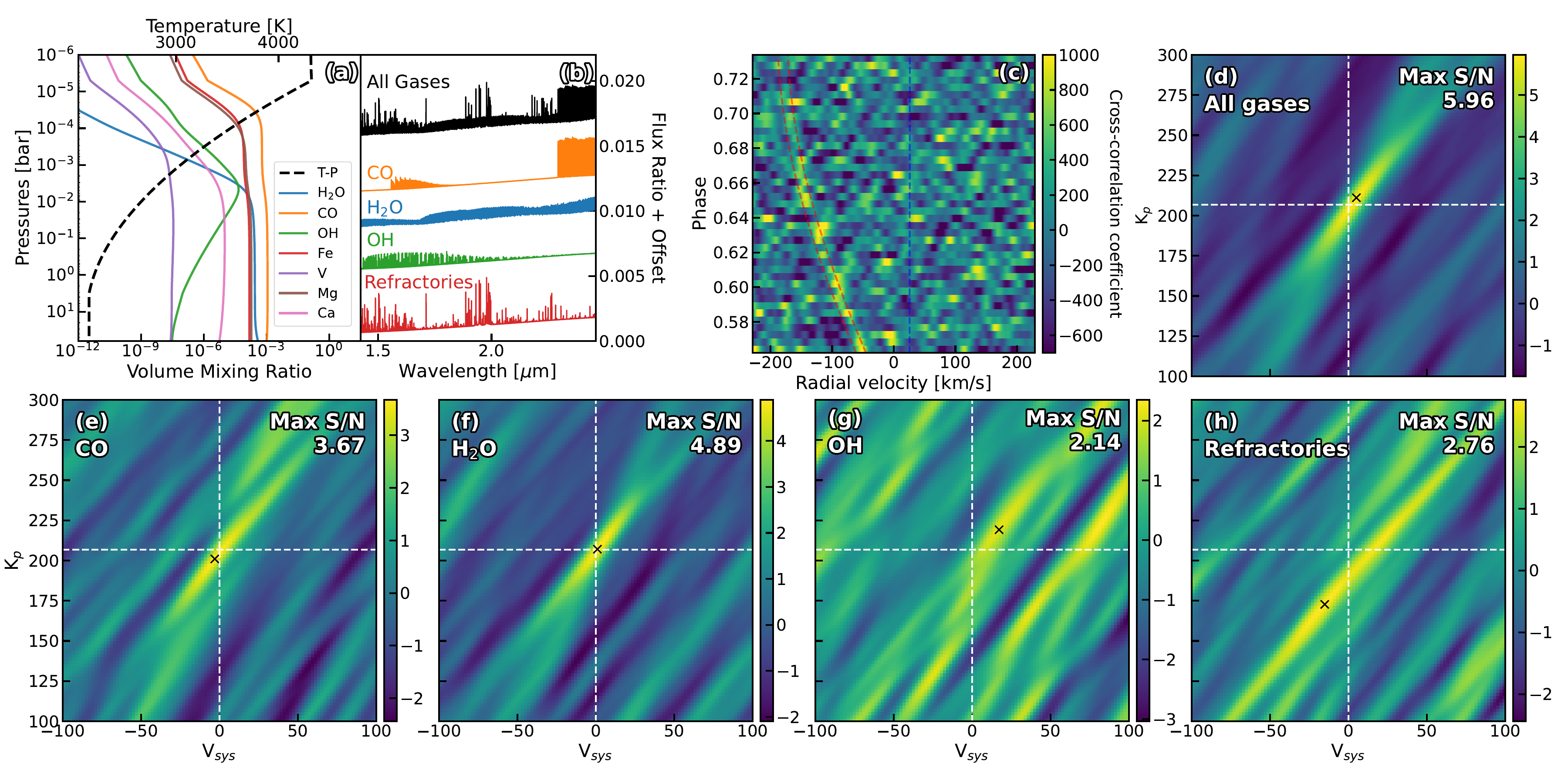}
\caption{\textbf{(a)}: Temperature--pressure profile (dashed black curve) and volume-mixing-ratio profiles of the selected opacity sources. \textbf{(b)}: Corresponding model planet-to-star flux ratio spectra over the IGRINS-2 wavelength range, showing the all-species spectrum, the individual volatile contributions from CO, H$_{2}$O, and OH, and the combined contribution from the refractory species. Each spectrum has a vertical offset for clarity. \textbf{(c)}: Cross-correlation trail for the all-species template as a function of orbital phase and radial velocity. The red dashed curve indicates the expected planetary radial velocity track. \textbf{(d--h)}: Integrated cross-correlation S/N maps in the $(v_{\mathrm{sys}}, K_{p})$ plane for the all-species template (\textbf{d}), the individual CO, H$_{2}$O, and OH templates (\textbf{e, f, g}, respectively), and the combined refractory template (\textbf{h}). White dashed lines indicate the expected planetary $v_{\mathrm{sys}}$ and $K_p$ values, and black crosses mark the position of the maximum S/N in each integrated map.}

\label{fig:detection}
\end{figure*}

We generated chemical-equilibrium abundance profiles with \texttt{FastChem} \citep{fastchem1,fastchem2} and high-resolution thermal emission spectra with \texttt{CHIMERA} \citep{chimera_paper,chimera_soft}, adopting the temperature--pressure parameterization of \citet{madsea09}. The abundance and temperature profiles are shown in Figure \ref{fig:detection}(a). The opacity set included CO, H$_2$O, OH, Ca, Mg, Fe, and V, together with H$_2$--H$_2$/H$_2$--He collision-induced absorption and H$^-$ bound-free and free-free opacity \citep{karman19}. Molecular cross-sections were generated from the HITEMP and POKAZATEL line lists \citep{rothman10,li15,polyansky18}, while the atomic data were adopted from \citet{kurucz18}. The spectra were initially calculated at $R\sim250{,}000$, convolved to the IGRINS-2 resolving power, rotationally broadened assuming tidally locked rotation ($v\sin i=9.1\,\mathrm{km\,s^{-1}}$), and converted to planet-to-star flux ratios using a \texttt{PHOENIX} stellar model \citep{phoenix} with $T_{\mathrm{eff}} = 6265\,\mathrm{K}$. We constructed an all-species template, individual templates for the volatile molecules, and a combined refractory template containing Ca, Mg, Fe, and V.

The planetary line-of-sight velocity at orbital phase $\phi$ was calculated as
\begin{equation}
v_{\rm LOS}(\phi)=v_{\rm bary}+v_{\rm sys}
+K_p\sin(2\pi\phi),
\end{equation}
where $v_{\rm bary}$ is the barycentric velocity, $K_p$ is the planetary orbital velocity amplitude, and $v_{\rm sys}$ is the residual velocity offset relative to the adopted system rest frame. To account for signal distortion introduced by detrending, each phase-shifted template was passed through the same SVD filtering applied to the observations. We then computed the cross-correlation function (CCF) following \citet{bro23} and integrated it along trial orbital tracks spanning $100\leq K_p\leq300\,\mathrm{km\,s^{-1}}$ and $-100\leq v_{\rm sys}\leq100\,\mathrm{km\,s^{-1}}$, with respective steps of $2\,\mathrm{km\,s^{-1}}$ in both parameters.
For each template, the integrated map was converted to S/N by subtracting its $3\sigma$-clipped median and dividing by its $3\sigma$-clipped standard deviation.
We applied the same filtering, cross-correlation, and normalization procedure to all templates, including the combined refractory template.
The resulting S/N is an empirical measure of the CCF signal strength and is not interpreted as a Gaussian equivalent statistical significance.

Figure~\ref{fig:detection} presents the template spectra, the abundance profile of each gas, the temperature--pressure profile used to generate the model, the phase-resolved CCF trail for the all-species template and the integrated CCF S/N maps.
The all-species template produces a peak of S/N$=5.96$ near the expected planetary velocity.
The individual CO and H$_2$O templates yield maximum S/N values of 3.67 and 4.89, respectively, and provide the primary evidence for the dayside atmospheric signal.
The OH template yields a weaker maximum S/N of 2.14.
The combined refractory map has a maximum S/N of 2.76, with the maximum located away from the expected planetary velocity.
Therefore, we regard the OH and combined refractory features as tentative.

\subsection{Atmospheric Retrieval}
\label{sec:retrieval}

\begin{table*}
\centering
\caption{Retrieval parameters, uniform priors, and posterior constraints. Posterior constraints are medians with 68\% credible intervals, except for $\log_{10}(P_2/\mathrm{bar})$ and $\log_{10}(P_3/\mathrm{bar})$, for which one-sided 95\% lower limits are reported.}
\label{tab:prior}
\begin{tabular}{llll}
\toprule
Parameter & Description & Prior & Posterior \\
\midrule
\multicolumn{4}{c}{\textit{Temperature Parameters}} \\
\midrule
$T_{0}$ & Reference temperature & $\mathcal{U}(500,\,6000)\,\mathrm{K}$ & $4315.26^{+1059.57}_{-925.09}\,\mathrm{K}$ \\
$\log_{10} (P_{1}/\mathrm{bar})$ & Pressure boundary & $\mathcal{U}(-10,\,2)$ & $-5.3^{+1.49}_{-1.58}$ \\
$\log_{10} (P_{2}/\mathrm{bar})$ & Pressure boundary & $\mathcal{U}(-5,\,1.9)$ & $>-0.005$ \\
$\log_{10} (P_{3}/\mathrm{bar})$ & Pressure boundary & $\mathcal{U}(-5,\,1.9)$ & $>-1.09$ \\
$\alpha_{1}$ & Temperature gradient parameter & $\mathcal{U}(0.02,\,1.0)$ & $0.61^{+0.26}_{-0.29}$ \\
$\alpha_{2}$ & Temperature gradient parameter & $\mathcal{U}(0.02,\,1.0)$ & $0.32^{+0.04}_{-0.04}$ \\

\midrule
\multicolumn{4}{c}{\textit{Abundance Parameters}} \\
\midrule
$\mathrm{[M/H]_{vol}}$ & Volatile element metallicity & $\mathcal{U}(-3,\,3)$ & $0.3^{+0.75}_{-0.59}$ \\
$\mathrm{[M/H]_{ref}}$ & Refractory element metallicity & $\mathcal{U}(-3,\,3)$ & $0.49^{+0.84}_{-0.68}$ \\
$\mathrm{[M/H]_{other}}$ & Other-element metallicity & $\mathcal{U}(-3,\,3)$ & $-0.78^{+1.67}_{-1.45}$ \\
$C/O$ & Carbon-to-oxygen number ratio & $\mathcal{U}(0.01,\,2.0)$ & $0.72^{+0.08}_{-0.10}$ \\

\midrule
\multicolumn{4}{c}{\textit{Dynamical Parameters}} \\
\midrule
$K_{p}$ & Planet orbital velocity amplitude & $\mathcal{U}(175,\,225)\,\mathrm{km\,s^{-1}}$ & $207.52^{+1.87}_{-1.96}\,\mathrm{km\,s^{-1}}$ \\
$v_{\rm sys}$ & Systemic velocity offset & $\mathcal{U}(-10,\,10)\,\mathrm{km\,s^{-1}}$ & $3.34^{+1.34}_{-1.43}\,\mathrm{km\,s^{-1}}$ \\
\bottomrule
\end{tabular}
\end{table*}

We performed a chemically consistent Bayesian retrieval to constrain the thermal structure and atmospheric composition.
The atmosphere was discretized into 80 layers uniformly spaced in log pressure from $10^{-6}$ to $10^2$ bar, with equilibrium chemistry used to account for non-uniform abundance profiles due to thermal dissociation, ionization, and recombination.
The composition was described by four parameters: $[\mathrm{M/H}]_{\rm vol}$ scales the total C+O inventory; C/O redistributes this inventory between carbon and oxygen; $[\mathrm{M/H}]_{\rm ref}$ jointly scales Ca, Mg, Fe, and V while fixing their relative abundances to solar values; and $[\mathrm{M/H}]_{\rm other}$ scales the remaining heavy elements included in \texttt{FastChem}.
These remaining elements were retained in the equilibrium chemistry calculations, but their direct line opacities were omitted because their contributions are negligible over the IGRINS-2 wavelength range.
The temperature structure was described by the six-parameter profile of \citet{madsea09}, and $K_p$ and $v_{\rm sys}$ were allowed to vary in the retrieval so that the model could localize the planetary signal self-consistently in velocity space.

For each sampled parameter set, \texttt{FastChem} calculated abundances under chemical equilibrium and \texttt{CHIMERA} generated the emission spectrum.
The model was Doppler-shifted to each observed phase and passed through the same detrending procedure as the data.
The likelihood was evaluated using the formulation of \citet{brogi19}.
We explored the 12-parameter model with \texttt{MultiNest} \citep{multinest,pymultinest}, adopting the uniform priors listed in Table~\ref{tab:prior}.

\begin{figure*}
\epsscale{1.1}
\vspace*{-0.1in}
\plotone{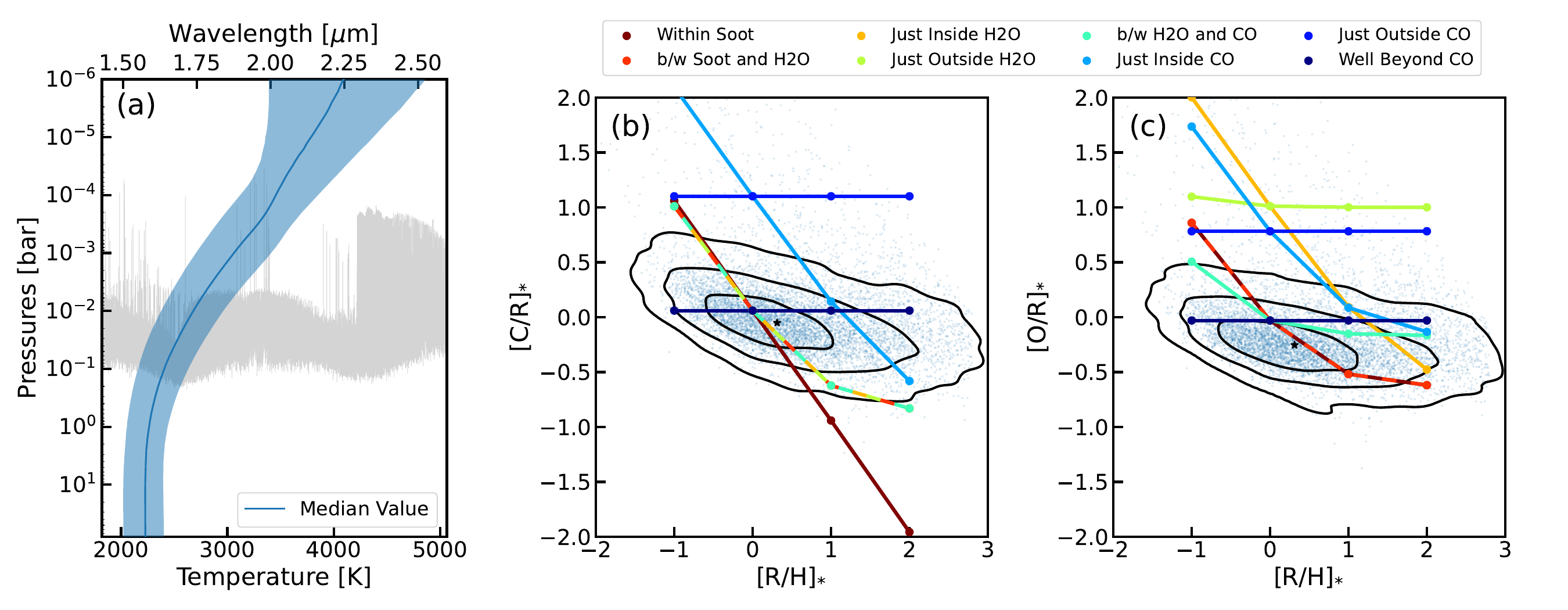}
\caption{\textbf{(a)}: Retrieved dayside temperature--pressure profile of WASP-12~b from the chemically consistent retrieval. The solid blue curve shows the median posterior profile, and the blue shaded region represents the $1\sigma$ credible range at each pressure level. The gray shading illustrates the wavelength-dependent pressure range probed by the emission. \textbf{(b)}: Comparison of the retrieved abundance-ratio posterior for WASP-12~b with the formation scenario predictions of \citet{chachan23} using [C/R]$_{*}$ versus [R/H]$_{*}$. \textbf{(c)}: [O/R]$_{*}$ versus [R/H]$_{*}$. Light-blue points show posterior samples, black contours indicate the 1, 2, and 3$\sigma$ regions of the posterior distribution, and the colored tracks correspond to the eight formation scenarios from \citet{chachan23}. Here, R denotes refractory elements, and all abundance ratios are shown relative to the host star composition.}
\label{fig:TPchachan}
\end{figure*}

The retrieved temperature--pressure profile, shown in Figure~\ref{fig:TPchachan} (a), is inverted over the pressure range probed by the IGRINS-2 data.
We obtain $[\mathrm{M/H}]_{\rm vol}=0.3^{+0.75}_{-0.59}$ and $[\mathrm{M/H}]_{\rm ref}=0.49^{+0.84}_{-0.68}$.
Both abundances have supersolar median values, although their credible intervals include solar abundances.
The retrieved C/O ratio is $0.72^{+0.08}_{-0.10}$.
The retrieval localizes the planetary signal at $K_{p}=207.52^{+1.87}_{-1.96}\,\mathrm{km\,s^{-1}}$ and $v_{\rm sys}=3.34^{+1.34}_{-1.43}\,\mathrm{km\,s^{-1}}$, consistent with the expected planetary signal.

\section{Discussion}
\label{sec:discussion}

\subsection{Atmospheric Composition and C/O}
\label{sec:ctoo}
\begin{figure}
\epsscale{1.1}
\plotone{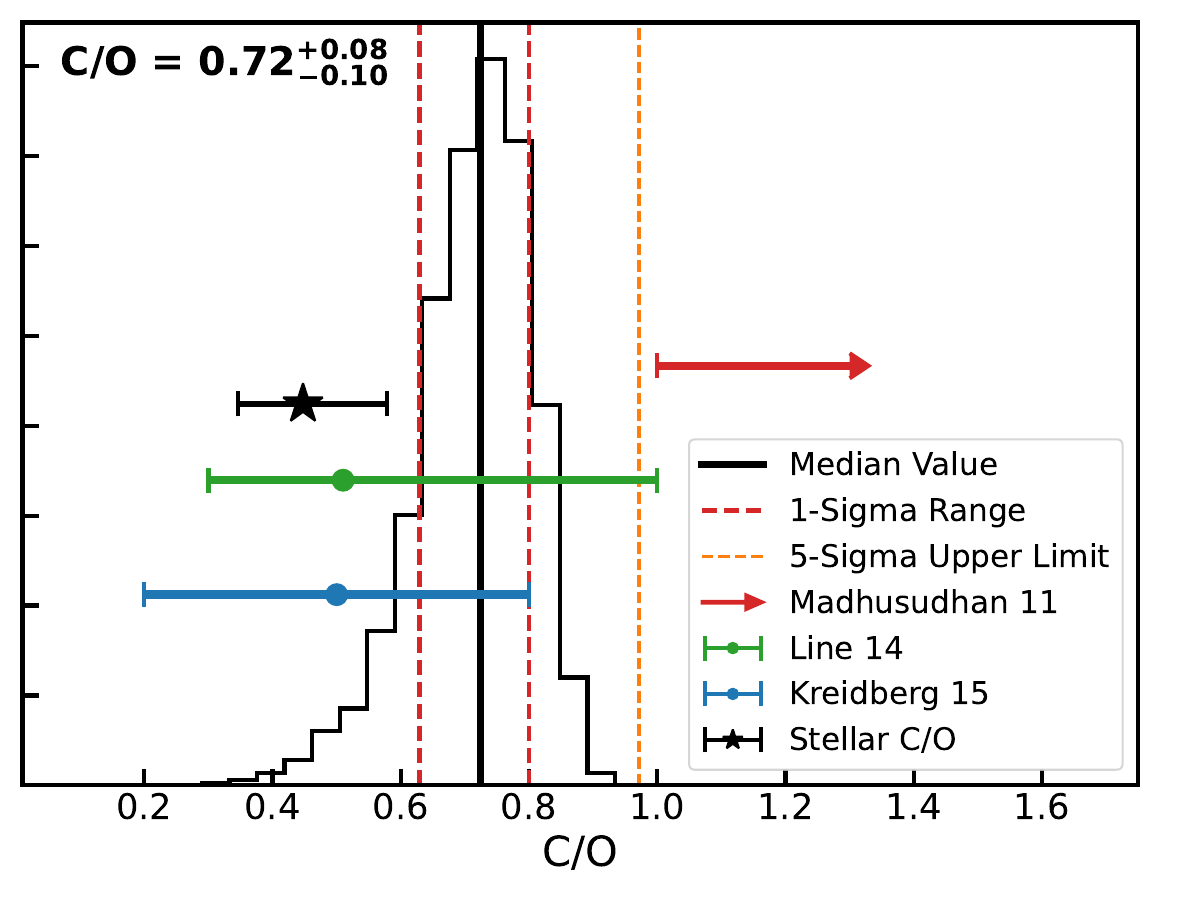}
\caption{Posterior distribution of the retrieved C/O ratio of WASP-12~b. The solid black, red dashed, and orange dashed lines respectively indicate the median, $1\sigma$ credible range, and $5\sigma$ upper credible limit of C/O=0.97. The black star marks the host star C/O ratio derived from the stellar abundances reported by \citet{fossati10}, with the $1\sigma$ error bar. The red arrow indicates the lower limit of C/O ratios from \citet{madhu11}, while the green and blue bars show the $1\sigma$ ranges from \citet{line14} and \citet{kreidberg15}.}
\label{fig:ctoo_post}
\end{figure}

Previous measurements of the atmospheric C/O ratio of WASP-12~b produced conflicting interpretations. Early analyses of its dayside emission favored a carbon-rich atmosphere with C/O $\geq 1$ \citep{madhu11,madhu12,stevenson14}, whereas \citet{line14} obtained a sub-unity solution. HST/WFC3 transmission spectroscopy subsequently favored C/O $< 1$ and rejected a carbon-rich terminator \citep{kreidberg15}, but this measurement does not necessarily imply C/O $< 1$ for the dayside atmosphere. More recently, the phase curve and emission spectrum retrieval of \citet{akinsanmi24} again favored mostly C/O $> 1$ solutions. Reanalyses of the earlier emission measurements demonstrated that the inferred dayside C/O was sensitive to the adopted eclipse depths, abundance priors, opacity data, and retrieval assumptions \citep{oreshenko17,himes22}. In particular, \citet{himes22} concluded that the available low-resolution dayside observations were insufficient to determine which C/O interpretation applied to WASP-12~b. 
Thus, despite strong evidence for an oxygen-rich terminator, whether the dayside atmosphere was carbon-rich remained unresolved.
However, no high-resolution spectroscopic observation had directly constrained the dayside C/O by resolving individual lines of CO, H$_2$O, and OH.
Therefore, our IGRINS-2 observations provide the missing line-resolved dayside constraint needed to address this controversy.

Our cross-correlation analysis recovers planetary signals in emission, including a strong H$_2$O signature, indicating a pronounced thermal inversion in the atmosphere of WASP-12~b.
In hot, hydrogen-dominated atmospheres under chemical equilibrium, C/O $>1$ strongly suppresses H$_{2}$O because most oxygen is bound in CO, whereas C/O $<1$ leaves excess oxygen available to form water \citep{madhu12}. Therefore, the observed H$_{2}$O emission directly supports an oxygen-rich dayside composition.
To quantify this interpretation while exploring both oxygen-rich and carbon-rich solutions, our chemically consistent retrieval adopts a uniform prior of C/O $\sim \mathcal{U}(0.01,2.0)$. Figure \ref{fig:ctoo_post} shows the retrieved posterior of C/O $=0.72^{+0.08}_{-0.10}$, with a $5\sigma$ upper credible limit of C/O $=0.97$. Therefore, carbon-rich solutions with C/O $>1$ are disfavored at greater than $5\sigma$ in our IGRINS-2 data. The retrieved value is higher than the near-solar central estimates of \citet{line14} and \citet{kreidberg15}, but remains consistent with their reported uncertainties. Our independent measurement provides a new dayside constraint and resolves the long-standing carbon-rich versus oxygen-rich ambiguity in favor of a mildly enhanced but sub-unity C/O.

\subsection{Formation Implications}
\label{sec:formation}
In our retrieval parameterization, the refractory metallicity, $[\mathrm{M/H}]_{\rm ref}$, scales the abundances of Ca, Mg, Fe, and V together at fixed solar relative abundances.
We simultaneously retrieve the volatile metallicity, $[\mathrm{M/H}]_{\rm vol}$, and the C/O ratio to constrain the relative volatile and refractory abundances.
Using these retrieved abundance parameters and the stellar abundances adopted from \citet{fossati10}, we calculate the carbon-to-refractory, oxygen-to-refractory, and refractory-to-hydrogen abundance ratios, $[\mathrm{C/R}]_{*}$, $[\mathrm{O/R}]_{*}$, and $[\mathrm{R/H}]_{*}$, respectively.
These quantities are expressed on a logarithmic scale relative to the corresponding host star abundance ratios.

Panels (b) and (c) of Figure~\ref{fig:TPchachan} compare our posteriors with the predicted composition tracks of \citet{chachan23}.
The $1\sigma$ credible regions overlap multiple prediction tracks in both abundance ratio planes, so the present measurements do not clearly favor a single formation scenario.
The tentative refractory signal likely contributes to the broad abundance constraints and hence to this limited discriminating power.
Also, our tentative refractory feature can be affected by cold-trapping which can deplete gas-phase refractory species relative to the planet's bulk composition.
Therefore, additional observations could improve the refractory abundance constraints and help distinguish among the competing formation scenarios.



\subsection{Comparison with Other IGRINS and IGRINS-2 Ultra-hot Jupiters}
\label{sec:population}
\begin{figure}
\epsscale{1.1}
\plotone{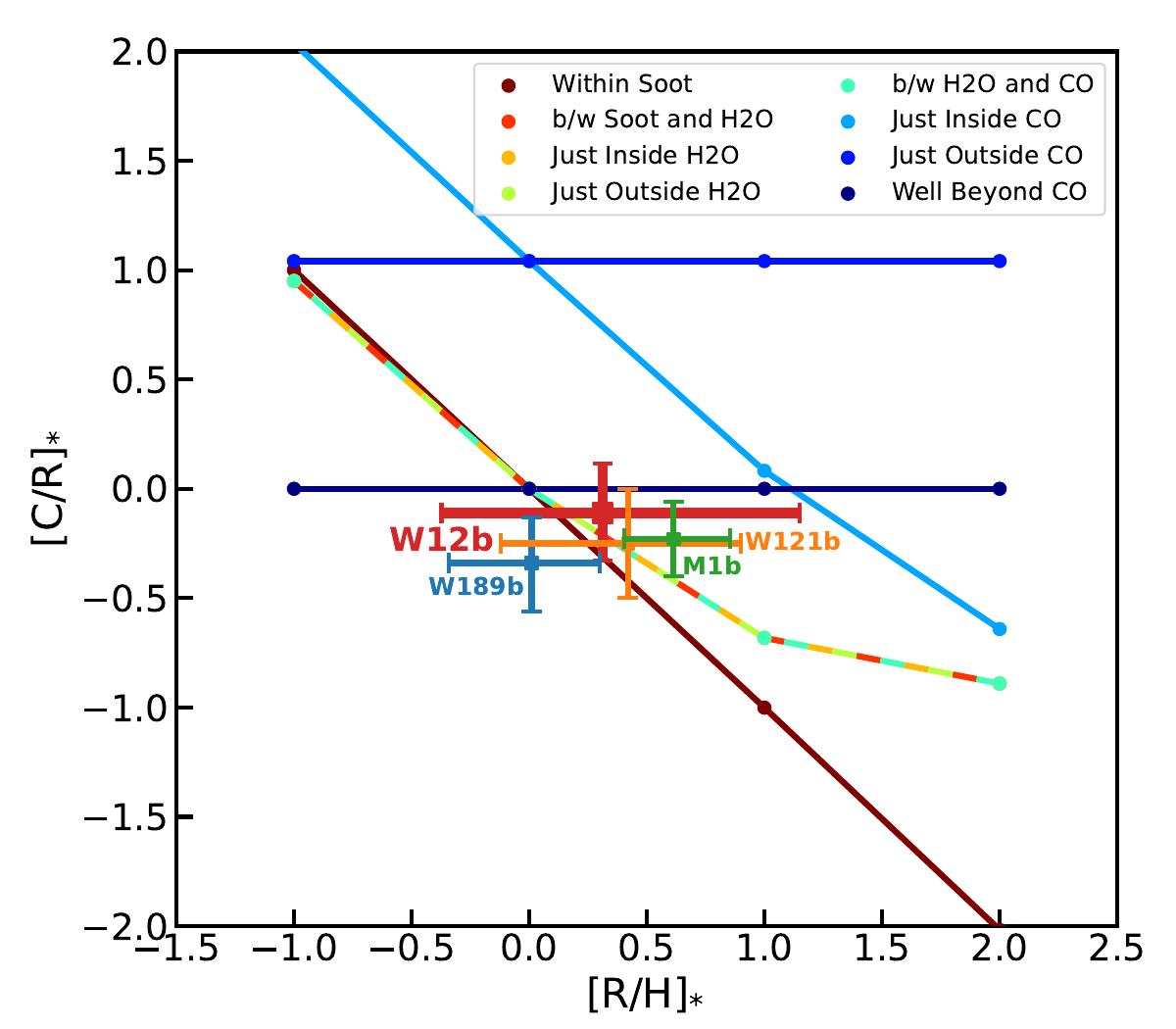}
\caption{Comparison of WASP-12~b with ultra-hot Jupiters observed using IGRINS or IGRINS-2 in the [C/R]$_*$--[R/H]$_*$ plane. Points and 1$\sigma$ error bars show the inferred host-star-relative abundance ratios for each planet, and the colored curves indicate the eight formation scenarios from \citet{chachan23}.}
\label{fig:prev_obs}
\end{figure}

We compare WASP-12~b with WASP-121~b, WASP-189~b, and MASCARA-1~b, whose volatile and refractory abundances have also been constrained using IGRINS or IGRINS-2 \citep{smith24b,sanchez26,kanumalla26}.
Figure~\ref{fig:prev_obs} presents this comparison in the generic $[\mathrm{C/R}]_{*}$--$[\mathrm{R/H}]_{*}$ plane, with R representing the refractory abundance scaling adopted in each study.
Although composition measurements from other instruments are available for some planets in our sample \citep[e.g.,][]{pelletier26}, we focus on IGRINS/IGRINS-2 results to compare abundances obtained with similar instrumentation.

The central values occupy a similar region, with stellar to superstellar refractory enrichments and substellar carbon-to-refractory ratios.
WASP-12~b has comparatively large uncertainties, and its $1\sigma$ intervals overlap those of each comparison planet in both $[\mathrm{R/H}]_{*}$ and $[\mathrm{C/R}]_{*}$.
Broad consistency is also seen in the reported refractory-to-volatile ratios.
WASP-12~b has $[\mathrm{R/V}]_{*}=0.17^{+0.22}_{-0.23}$, compared with $0.58^{+0.29}_{-0.25}$ for WASP-121~b \citep{smith24b}, $0.38^{+0.18}_{-0.19}$ for WASP-189~b \citep{sanchez26}, and $[\mathrm{R/V}]_{\odot}=0.36^{+0.11}_{-0.09}$ for MASCARA-1~b \citep{kanumalla26}.
For MASCARA-1~b, the refractory-to-volatile ratio is reported relative to solar abundances because the host star oxygen abundance has not been measured.
Because of the tentative refractory signal and broad abundance constraints for WASP-12~b, the apparent consistency should be interpreted cautiously.
Despite these uncertainties, the consistency across abundance ratios suggests a possible common pattern in the atmospheric elemental abundances of these planets.
Such similarities could reflect comparable chemical processing and similar relative contributions from volatile-rich gas and refractory-rich solids during formation.
Interestingly, these systems also exhibit substantial sky-projected spin--orbit misalignments: $\lambda=59^{\circ}\,^{+15^{\circ}}_{-20^{\circ}}$ for WASP-12~b \citep{albrecht12}; $\lambda=102.2^{\circ}\,^{+5.5^{\circ}}_{-5.3^{\circ}}$ for WASP-121~b \citep{delrez16}; $\lambda=89.3^{\circ}\pm1.4^{\circ}$ for WASP-189~b \citep{anderson18}; and $\lambda=69.5^{\circ}\pm3.0^{\circ}$ for MASCARA-1~b \citep{talens17}, raising the possibility that their histories share common features.
Therefore, the apparent compositional and orbital similarities motivate further investigation of whether common formation and evolutionary processes operated across these planets.

Previous R/V studies illustrate that atmospheric abundance measurements do not always yield unique formation locations.
For WASP-121~b, \citet{smith24b} favor accretion between the soot line and the H$_2$O snowline while allowing more distant locations, whereas \citet{pelletier25} infer accretion near the CO snowline.
Multiple formation regions also remain compatible with the results for WASP-189~b and MASCARA-1~b \citep{sanchez26,kanumalla26}.
Therefore, similar atmospheric abundance ratios across planets need not imply a common formation history.
High-resolution retrievals also often constrain relative abundance ratios more precisely than absolute enrichments \citep{gibson22}, and these differing measurement precisions should be considered when assessing apparent population patterns.
In addition to these limitations, our comparison may be affected by selection and analysis biases arising from the small sample size and analyses which are not fully homogeneous, even though the present comparison uses similar instrumentation and comparable analyses.
Expanding this sample and applying a common analysis framework would enable consistent statistical comparisons of elemental abundances while accounting for measurement uncertainties.
Such population studies will allow us to test whether the apparent chemical similarities persist and to place broader constraints on the processes shaping giant planet compositions.

\section{Conclusion}
\label{sec:conclusion}
We presented high-resolution near-infrared dayside observations of WASP-12~b obtained with Gemini North/IGRINS-2.
Cross-correlation analysis recovered emission signals from H$_2$O and CO, together with tentative signals from OH and gas-phase refractory species.
Our chemically consistent retrieval favors an inverted thermal structure and yields C/O $=0.72^{+0.08}_{-0.10}$ and $[\mathrm{R/V}]_{*}=0.17^{+0.22}_{-0.23}$.
The volatile and refractory enrichments have supersolar median values, although both remain consistent with solar abundances within their 1$\sigma$ credible ranges.
Our independent and line-resolved measurement disfavors carbon-rich dayside atmospheres with C/O $>1$ at greater than 5$\sigma$ and strengthens the evidence for an oxygen-rich atmosphere suggested by \citet{kreidberg15}.

Using the retrieved volatile and refractory abundances, we compared WASP-12~b with other ultra-hot Jupiters characterized using IGRINS/IGRINS-2.
The $1\sigma$ credible intervals for WASP-12~b overlap those of the comparison planets in both $[\mathrm{C/R}]_{*}$ and $[\mathrm{R/H}]_{*}$, while the reported refractory-to-volatile ratios are broadly consistent.
These similarities suggest a possible shared atmospheric abundance pattern, although the tentative refractory signal of WASP-12~b and small sample size limit the strength of this interpretation.
Additional observations could improve the refractory abundance constraints, while a larger sample analyzed within a common retrieval framework would enable statistical tests of these apparent chemical similarities and their possible implications for giant planet formation and evolution.

\begin{acknowledgments}
This work was supported by K-GMT Science Program (PID: GN-2025A-Q-112) of Korea Astronomy and Space Science Institute (KASI).
Based on observations obtained at the international Gemini Observatory, a program of NSF NOIRLab, which is managed by the Association of Universities for Research in Astronomy (AURA) under a cooperative agreement with the U.S. National Science Foundation on behalf of the Gemini Observatory partnership: the U.S. National Science Foundation (United States), National Research Council (Canada), Agencia Nacional de Investigaci\'{o}n y Desarrollo (Chile), Ministerio de Ciencia, Tecnolog\'{i}a e Innovaci\'{o}n (Argentina), Minist\'{e}rio da Ci\^{e}ncia, Tecnologia, Inova\c{c}\~{o}es e Comunica\c{c}\~{o}es (Brazil), and Korea Astronomy and Space Science Institute (Republic of Korea).
S.P.\ is funded by the Courtois Scientific Vanguard Fund of the Courtois Foundation and acknowledges support from the Trottier Institute for Research on Exoplanets. This work was supported by grant https://doi.org/10.69777/377645 from the Fonds de recherche du Québec and by the Center for research in astrophysics of Québec (AstroQuébec).
\end{acknowledgments}





%
\facilities{Gemini:North(IGRINS-2)}


\software{\textsc{pymultinest} \citep{pymultinest}, \textsc{astropy} \citep{astropy}, \textsc{numpy} \citep{numpy}, \textsc{scipy} \citep{scipy}, \textsc{matplotlib} \citep{matplotlib}, \textsc{fastchem} \citep{fastchem1, fastchem2}
          }


\appendix

\setcounter{figure}{0}
\renewcommand\thefigure{A\arabic{figure}}
\section{SVD Optimization}
\label{appendix:SVD}

\begin{figure*}
\plotone{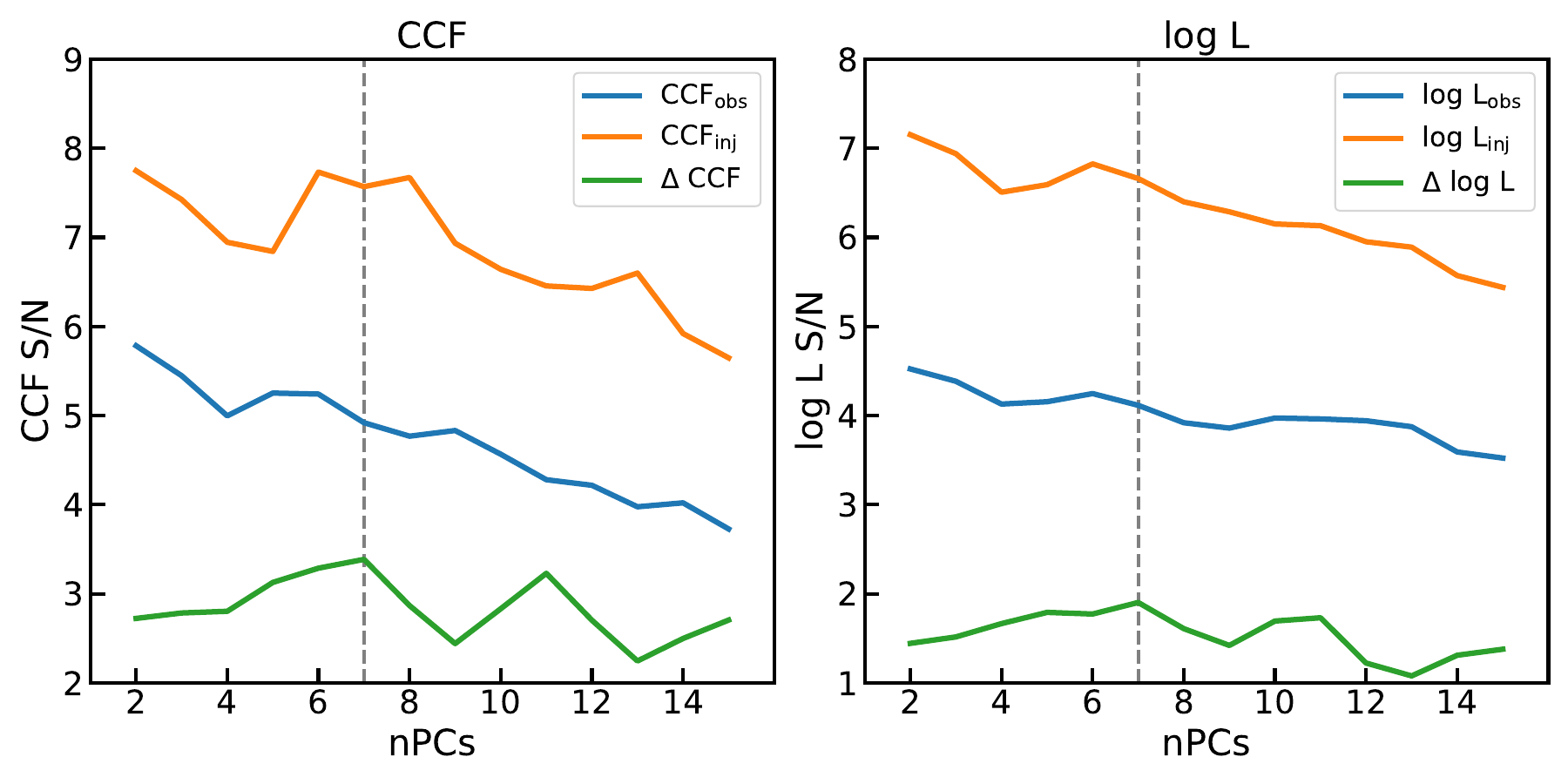}
\caption{Optimization of the number of removed SVD components following \citet{cheverall23}. The left and right panels show the S/N curves from the CCF and log-likelihood values, respectively. Blue, orange, and green curves indicate the observed, signal-injected, and differential results, respectively. Gray dashed lines mark the adopted value of seven components, which maximizes both differential metrics.}
\label{fig}
\label{fig:svd_opt}
\end{figure*}

We optimized the number of removed SVD components using the differential optimization method of \citet{cheverall23}. 
A solar-metallicity model containing the same opacity sources as the main analysis was injected into the observed spectra at the expected planetary velocity and was also used as the cross-correlation template in the optimization process.
We tested the removal of two to fifteen components and calculated $\Delta\mathrm{CCF}=\mathrm{CCF}_{\rm inj}-\mathrm{CCF}_{\rm obs}$ for each case, where $\mathrm{CCF}_{\rm inj}$ and $\mathrm{CCF}_{\rm obs}$ denote the cross-correlation functions obtained with and without the injected model, respectively.
Figure~\ref{fig:svd_opt} shows the resulting S/N curves for the observed, signal-injected, and differential CCF and log-likelihood values.
The $\Delta\mathrm{CCF}$ S/N reaches its maximum at seven components, and the differential log-likelihood metric favors the same choice. 
Therefore, we adopted seven components for every spectral order in the subsequent cross-correlation and retrieval analyses.

\section{Retrieval Corner Plot}
\label{appendix:corner}

\begin{figure}[ht!] 
\centering
\includegraphics[width=\textwidth]{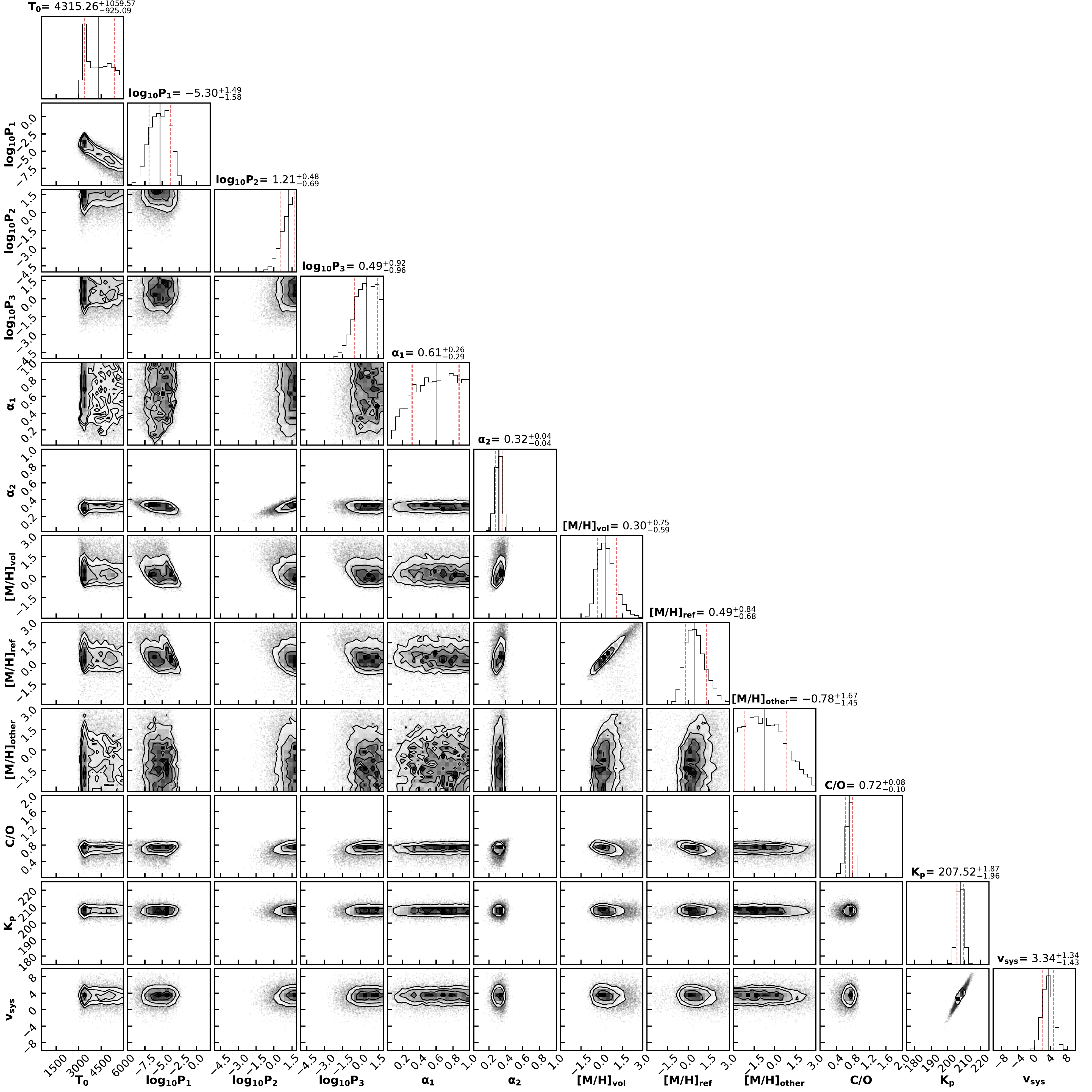}
\caption{
Corner plot showing the marginalized posterior distributions of the retrieval parameters. The retrieved parameters describe the pressure–temperature profile, the metallicities of the volatile, refractory, and other elements, and the velocity parameters ($K_{p}$, $v_{\rm sys}$). 
}
\label{fig:corner}
\end{figure}

Figure~\ref{fig:corner} presents the full joint posterior distribution from our chemically consistent retrieval. The parameter definitions and adopted uniform priors are listed in Table~\ref{tab:prior}. The diagonal panels show the marginalized posterior distributions, while the off-diagonal panels show the joint posterior distributions for pairs of retrieved parameters.

\section{log-likelihood map}
\label{appendix:logL}

\begin{figure*}
\plotone{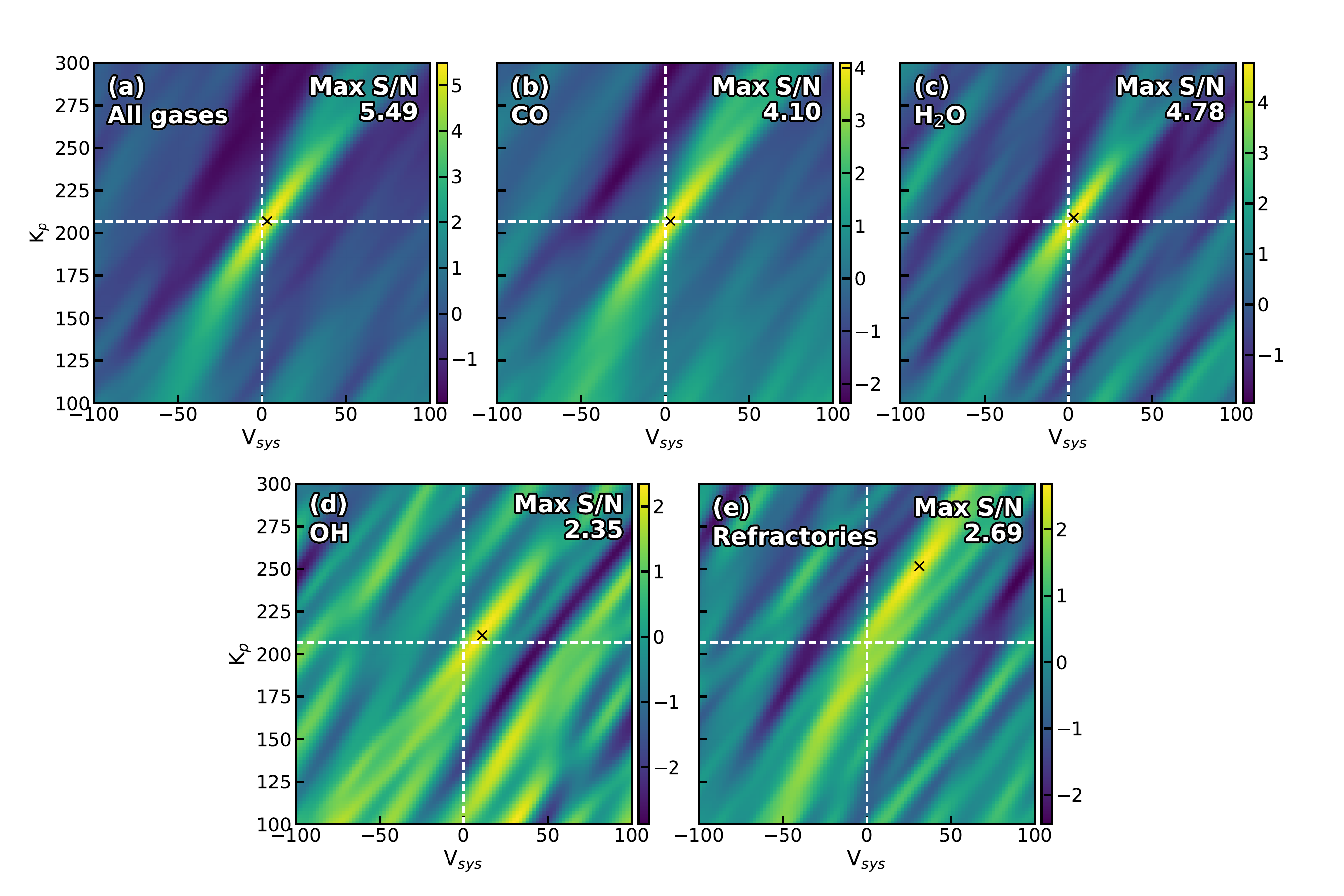}
\caption{\textbf{(a--e)}: Log-likelihood maps in the $(v_{\mathrm{sys}}, K_{p})$ plane for the all-species template (\textbf{a}) and the individual CO, H$_2$O, and OH templates (\textbf{b, c, d}, respectively), and the combined refractory template (\textbf{e}). White dashed lines indicate the expected planetary $v_{\mathrm{sys}}$ and $K_{p}$ values, and black crosses mark the position of the maximum value in each map.}
\label{fig}
\label{fig:logL}
\end{figure*}

Figure~\ref{fig:logL} presents the log-likelihood maps corresponding to the templates examined in the cross-correlation analysis. These maps provide a complementary assessment of the planetary signal in the $(v_{\mathrm{sys}}, K_p)$ plane.
These log-likelihood maps are calculated using the likelihood formulation of \citet{brogi19} and the same orbital-track integration and S/N normalization as the CCF maps in Section~\ref{sec:method}.

\section{Injection--Recovery Test}
\label{appendix:injection}

\begin{figure*}
\plotone{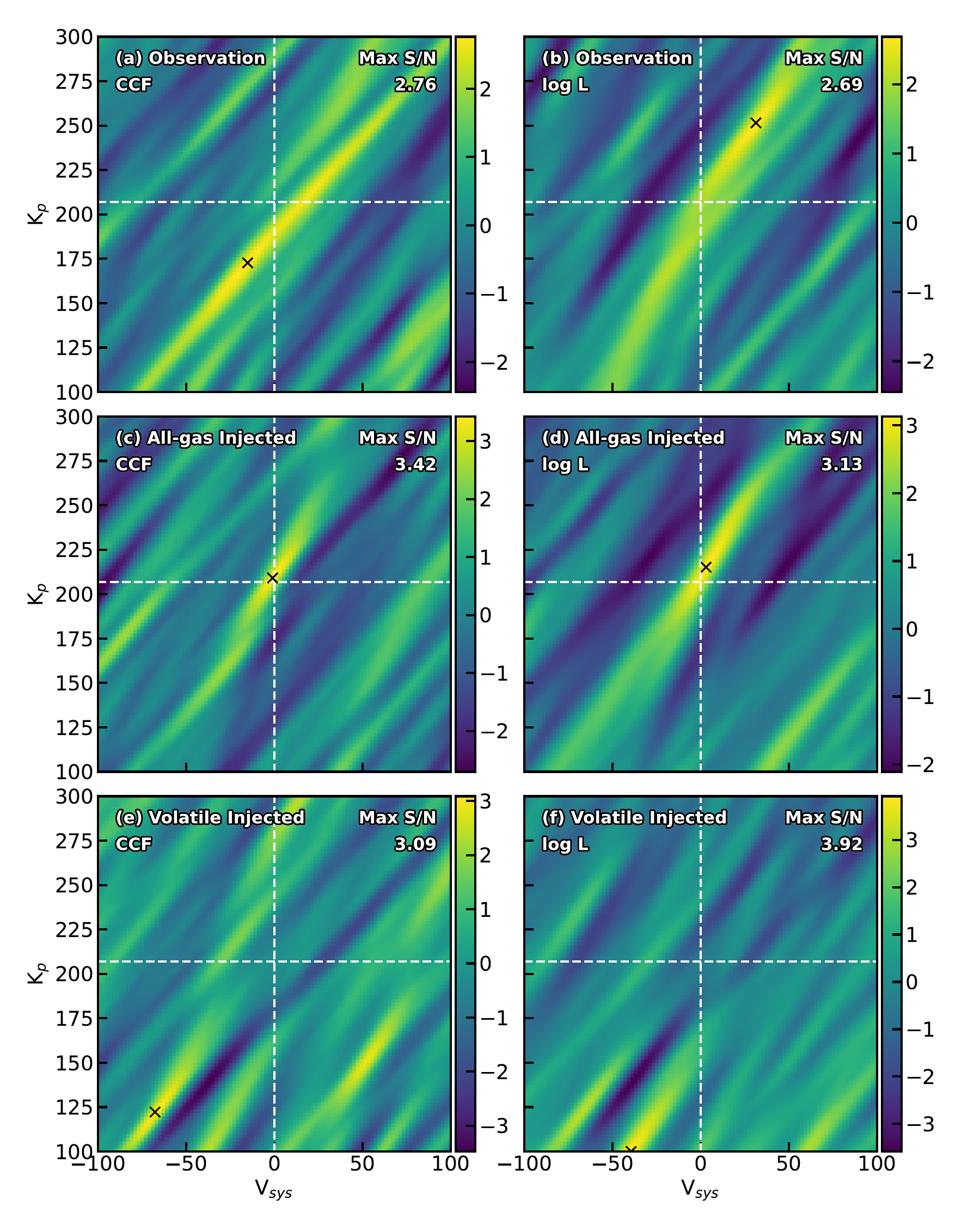}
\caption{Injection--recovery tests for the bulk refractory signal. The left and right columns show the CCF S/N and log-likelihood S/N maps, respectively, calculated using the combined refractory template. \textbf{(a, b)}: Maps obtained from the observations. \textbf{(c, d)}: Maps obtained from a synthetic data set containing the injected all-species template with the retrieved abundances. \textbf{(e, f)}: Maps obtained from a synthetic data set containing only the injected volatile signal. White dashed lines indicate the expected planetary $v_{\mathrm{sys}}$ and $K_p$ values, and black crosses mark the position of the maximum value in each map.}
\label{fig:InjRec}
\end{figure*}

We performed injection–recovery tests to assess whether volatile line cross-talk could reproduce the tentative refractory feature and whether a refractory signal with the retrieved abundances could be recovered.
Synthetic data sets were constructed using the best-fitting all-species spectrum and its volatile-only counterpart, with the volatile and refractory abundances in the former set by the retrieved $[\mathrm{M/H}]_{\rm vol}$ and $[\mathrm{M/H}]_{\rm ref}$.
Each spectrum was Doppler-shifted to the observed phases, combined with the stellar and telluric components reconstructed from the SVD decomposition, and subjected to S/N-dependent Poisson noise.
Both synthetic data sets were processed through the same detrending pipeline as the observations and analyzed using the combined refractory template containing Ca, Mg, Fe, and V.
Figure~\ref{fig:InjRec} compares the resulting CCF and log-likelihood maps with those obtained from the observations.
The volatile-only injection does not reproduce a comparable refractory feature near the expected planetary velocity, whereas the all-species injection produces a feature near the injected velocity in both maps.
These results suggest that volatile line cross-talk alone is unlikely to produce the observed feature and the refractory signal can be recovered with the retrieved abundances.
These results support the interpretation of the observed feature as a refractory contribution, although its low S/N warrants retaining its tentative classification.

\bibliography{sample702}{}
\bibliographystyle{aasjournalv7.1}



\end{document}